\documentclass{elsart3p}
\usepackage{times,mathptmx}
\usepackage[numbers,sort&compress]{natbib}
\usepackage{amssymb}
\usepackage[final]{graphicx}
\usepackage{units}
\usepackage{amsmath}
\def\lsi{\raise0.3ex\hbox{$<$\kern-0.75em\raise-1.1ex\hbox{$\sim$}}}
\def\gsi{\raise0.3ex\hbox{$>$\kern-0.75em\raise-1.1ex\hbox{$\sim$}}}
\newcommand{\lsim}{\mathop{\lsi}}

\def\be{\begin{equation}}
\def\ee{\end{equation}}
\def\ba{\begin{eqnarray}}
\def\ea{\end{eqnarray}}

\begin{document}

%%%%%%%%%%%%%%%%%%%%%%%%%%%%%%%%%%%%%%%%%%%%%%%%%%%%%%%%%%%%%%%%%%%%%%

\begin{frontmatter}
\title{Scale invariance, unimodular gravity and dark energy}

\author{Mikhail Shaposhnikov}
\ead{Mikhail.Shaposhnikov@epfl.ch} and
\author{Daniel Zenh\"ausern}
\ead{Daniel.Zenhaeusern@epfl.ch}

\address{
  Institut de Th\'eorie des Ph\'enom\`enes Physiques,
  \'Ecole Polytechnique F\'ed\'erale de Lausanne,
  CH-1015 Lausanne, Switzerland}

\date{14 November 2008}

\begin{abstract}
We demonstrate that the combination of the ideas of unimodular
gravity, scale invariance, and the existence of an exactly massless
dilaton leads to the evolution of the universe supported by present
observations: inflation in the past, followed by the radiation and
matter dominated stages and accelerated expansion at present. All
mass scales in this type of theories come from one and the same
source. 

\end{abstract}


\end{frontmatter}

%%%%%%%%%%%%%%%%%%%%%%%%%%%%%%%%%%%%%%%%%%%%%%%%%%%%%%%%%%%%%%%%%%%%%%
\section{Introduction}
\label{sec:intro}
The origin of different mass scales in particle physics is a mystery.
The masses of quarks, leptons and intermediate vector bosons come from
the vacuum expectation value (vev) of the Higgs field; the
dimensionful parameters like the QCD scale $\Lambda_{QCD}$ or the
scales related to
the running of all other dimensionless couplings of the Standard Model
(SM) are believed to have nothing to do with the Higgs vev. Newton's
gravitational constant provides yet another mass scale, very different
from typical particle masses of the SM. The Higgs mass itself -- 
where does it come from?

Is it possible that {\em all} these mass scales originate from one and
the same source?\footnote{We will refer to this possibility as
``no-scale scenario''.} Indeed, it is not difficult to construct, on
the {\em classical level}, a theory containing a new singlet field
$\chi$, which gives masses to all particles and fixes Newton's
constant. Having in mind the SM extended by 3 light right-handed
singlet fermions, the $\nu$MSM of \cite{Asaka:2005an,Asaka:2005pn}
(this theory -- Neutrino Minimal SM -- unlike the SM, can explain
neutrino masses and oscillations, dark matter and baryon asymmetry of
the universe), one can write the Lagrangian realizing this idea in the
following form\footnote{This expression is identical to the one in
\cite{Shaposhnikov:2008pf}, Section 8, but uses different notations.}
\begin{eqnarray}
\nonumber
  {\cal L}_{\nu \rm{MSM}}= {\cal L}_{\rm{SM}[M\rightarrow 0]} 
 + {\cal L}_G  + \frac{1}{2}(\partial_\mu\chi)^2 - V(\varphi,\chi)\\
   +  \left(\bar{N_I} i \gamma^\mu\partial_\mu N_I 
   - h_{\alpha I} \,  \bar L_\alpha N_I \tilde{\varphi} 
    - f_I \bar {N_I}^c N_I \chi + \rm{h.c.}\right) 
   \,,
  \label{lagr}
\end{eqnarray}
where the first term is the SM Lagrangian without the Higgs potential,
$N_I$ (I=1,2,3) are the right-handed singlet leptons, $\varphi$ and
$L_\alpha$ ($\alpha=e,\mu,\tau$) are  the Higgs and lepton doublets
respectively, $h_{\alpha I}$ and $f_I$ are the matrices of Yukawa
coupling constants. The scalar potential is given by
\be
V(\varphi,\chi) =
\lambda\left(\varphi^\dagger\varphi-\frac{\alpha}{2\lambda} \chi^2\right)^2+
\beta(\chi^2-\chi_0^2)^2~,
\label{pot}
\ee
and the gravity part is
\be
{\cal L}_G = -\left(\xi_\chi \chi^2 
+2 \xi_h\varphi^\dagger\varphi\right)\frac{R}{2}~,
\label{GG}
\ee 
where $R$ is the scalar curvature. We will only consider positive
values for $\xi_\chi$ and $\xi_h$, for which the coefficient in front
of the scalar curvature is positive, whatever values the scalar fields
take. This is the Lagrangian of ``induced gravity'' going back to
Refs. \cite{Zee:1978wi,Smolin:1979uz} (see also
\cite{Shaposhnikov:2006xi,Anisimov:2008qs} in the $\nu$MSM context).

For positive $\lambda$ and $\beta$ the theory with potential
\eqref{pot} possesses a ground state\footnote{By a
ground state we mean a constant solution of the equations of motion
including gravity. The existence of such a ground state could be
essential for a consistent quantization of the theory.}. It 
corresponds to the fields sitting at the minimum of the potential,
i.e. $\chi=\chi_0$, $h=h_0$ with ${h_0}^2=\frac{\alpha}{\lambda}
\chi_0^2$ and a constant metric describing flat space-time. The field
values at the potential minimum can be related to the Planck scale as
${M_P^2=\xi_\chi \chi_0^2 + \xi_h {h_0}^2}$, ${M_P= 2.44\times
10^{18}}$
GeV. Physics in this theory does not depend on a specific value of
$\chi_0$ -- all dimensionful parameters are proportional to it -- and
only dimensionless ratios can be measured.

Although the aim of having one source for all mass scales is achieved
by construction of the Lagrangian (\ref{lagr}) (we stress that we are
still discussing the classical theory) the solution is not
satisfactory: the absence of explicit mass terms for the Higgs field
and for singlet fermions, and the absence of a gravity scale along
with the introduction of the dimensionful parameter $\chi_0$, 
required to realize the scenario, are ad hoc and do not follow from
any symmetry principle.

The symmetry that forbids (on the classical level) the appearance of
any dimensionful parameters is well known --  it is the dilatational
symmetry. Under dilatations, the scalar and fermionic fields change as
${\phi(x)\to\sigma^n\phi(\sigma x)}$ ($n=1$ for scalars and ${n=3/2}$
for fermions), while the metric transforms as  ${g_{\mu\nu}(x)\to
g_{\mu\nu}(\sigma x)}$. The action (\ref{lagr}) is invariant under
this
symmetry, provided $\chi_0 = 0$, leading to the absence of all
dimensionful parameters. 

From now on we will require that a dilatation invariant theory should
possess a ground state. Since this requirement is essential
for our model, we will further discuss it in section
\ref{sec:dreams}. For the dilatation invariant theory to contain
massive singlet and doublet fermions, the ground state
should be such that $\chi\neq 0$ and $h\neq 0$. The only way to
achieve this is to set $\beta=0$. Thus, the no-scale scenario can only
be realized if $\beta=0$. In this case the potential (\ref{pot})
acquires the flat direction  ${h^2-\frac{\alpha}{\lambda} \chi^2=0}$,
and the theory contains one exactly massless particle $\eta$ -- a
certain mixture of the singlet $\chi$ and the Higgs field. The
requirement $\beta=0$ therefore leads to a theory with spontaneously
broken scale invariance, where $\eta$ appears as a Goldstone
boson\footnote{Spontaneous breaking of the dilatational symmetry also
occurs if the potential has a flat direction either along $\chi=0$ or
along $h=0$. However, both cases are unsatisfactory. The first one
corresponds to a theory with no massive singlet fermions, whereas the
second one is a theory with no electroweak symmetry breaking.}.
Equivalent arguments were given in \cite{Buchmuller:1988cj}.

The theory (\ref{lagr}) with $\beta=0$ (from now on only this choice
of parameters will be considered) is rather peculiar\footnote{We
stress that this theory is {\em not invariant} under local conformal
transformations. Conformal invariance requires the specific values
for $\xi_\chi$ and $\xi_h$, $\xi_\chi=\xi_h=-\frac{1}{6}$.}: not only
is the physics independent of the value of $\chi \neq 0$ in the ground
state, but the ground state is infinitely degenerate.  The question
``Who gives the mass to the dilaton ?'' does not arise. It is
massless, and the chain  of questions ``Who gives mass to whom?''
terminates. 

Not to any surprise, the classical scale-invariant theory constructed
in this way does not contain a cosmological constant $\Lambda$. So, if
we confront it with cosmological observations, it seems to fail, since
the universe is in accelerated expansion, which requires the presence
of dark energy with an equation of state close to that of the
cosmological constant. This conclusion is certainly correct for 
standard General Relativity (GR), associated with the action 
\be 
S_E = \int \sqrt{-g} d^4x {\cal L}_{\nu \rm{MSM}}~,
\label{action}
\ee
where $g$ is the determinant of the metric.

The aim of this Letter is to show that the situation is completely
different if general relativity in (\ref{action}) is replaced by
Unimodular Gravity (UG). UG is a very modest modification of
Einstein's theory: it adds a constraint $g=-1$ to the action principle
defined by eq. (\ref{action})
\cite{vanderBij:1981ym,Wilczek:1983as,Zee:1983jg,
Buchmuller:1988wx,Weinberg:1988cp,
Unruh:1988in,Alvarez:2005iy,Henneaux:1989zc}. UG is invariant under
diffeomorphisms which conserve the 4-dimensional volume element. It
contains the same number of dynamical degrees of freedom (massless
graviton) as Einstein's theory. To the best of our knowledge, the
consequences of scale-invariant UG with massless dilaton have not been
considered previously.

The relevance of UG for the cosmological constant problem was
realized long time ago
\cite{Wilczek:1983as,Zee:1983jg,Buchmuller:1988wx,
Weinberg:1988cp,vanderBij:1981ym}. If $g=-1$, adding a constant
$\Lambda$ to the Einstein-Hilbert action does not change the equations
of motion. Still, the $\Lambda$ problem is not solved, since the
cosmological constant shows up again, but now as an initial condition
for cosmological evolution in UG. We will see that for our case of a
scale-invariant theory together with UG, the initial conditions lead
to a non-trivial run-away effective potential for the dilaton rather
than to a cosmological constant, and thus to dynamical dark energy. 
Moreover, it will turn out that both inflation and accelerated
expansion of the universe can be explained on the same footing.

The Letter is organized as follows. In Section \ref{sec:unimod} we
show
that scale-invariant UG with a massless dilaton is equivalent (on the
classical level) to Einstein's theory with zero cosmological constant
and a peculiar potential, the magnitude of which is fixed by initial
conditions for all the fields. We continue in Section \ref{sec:dark}
with a
discussion of the evolution of the universe in our model.  The
requirements a full quantum theory should satisfy for our findings to
remain valid are formulated in Section \ref{sec:dreams}. Section
\ref{sec:concl} is a summary of the results.

%%%%%%%%%%%%%%%%%%%%%%%%%%%%%%%%%%%%%%%%%%%%%%%%%%%%%%%%%%%%%%%%%%%%%%
\section{Scale-invariant unimodular gravity : the classical theory}
\label{sec:unimod}
In this Letter we want to bring together several a priori separate
ideas. One of them is unimodular gravity, which has appeared many
times in the literature
\cite{vanderBij:1981ym,Wilczek:1983as,Zee:1983jg, Weinberg:1988cp,
Unruh:1988in,Alvarez:2005iy,Henneaux:1989zc}. In unimodular gravity
one reduces the dynamical components of the metric $g_{\mu\nu}$ by
one, imposing that the metric determinant $g\equiv\det(g_{\mu\nu})$
takes some fixed constant value.\footnote{In principle one can fix
$g=a(x)$, where $a(x)$ is a fixed external field, and the results are
the same.} Conventionally one takes ${|g|=1}$, hence the name. Fixing
the metric determinant to one is not a strong restriction, in the
sense that the family of metrics satisfying this requirement can still
describe all possible geometries. For pure gravity, things are very
simple and well known. The analog of the Einstein-Hilbert Lagrangian
for unimodular gravity is
\begin{equation}
 \mathcal{L}_{EH}=-M_P^2\,\frac{\hat{R}}{2}\;.
\end{equation} 
Writing quantities with a hat, like $\hat{R}$, we mean that they
depend on the metric with $g=-1$. These quantities transform like
tensors under the group of volume preserving diffeomorphisms, i.e.
coordinate transformations $x^\mu\rightarrow\xi^\mu(x)$, with the
condition $\hat{\nabla}_\mu\xi^\mu=0$.\footnote{It is important to
distinguish UG from theories constructed on the simple requirement of
invariance under restricted coordinate transformations
$x^\mu\rightarrow\xi^\mu(x)$, with $\hat{\nabla}_\mu\xi^\mu=0$
(sometimes called TDiff gravity)
\cite{Buchmuller:1988wx,Pirogov:2006zd,Alvarez:2006uu}. The latter
theories contain in general a third dynamical degree of freedom for
the metric. In addition, they have field equations, which depend on
the choice of coordinates. In UG the constraint on the metric
determinant is essential. It is responsible for the absence of a third
metric degree of freedom and guarantees that the equations of motion
do not depend on the coordinate choice.} Doing variations of this
action that keep the metric determinant fixed, since it is not a
dynamical variable, yields the equations of motion
\begin{equation}
 \hat{G}_{\mu\nu}=-\Lambda\, \hat{g}_{\mu\nu}\;,
\end{equation}
where $\Lambda$ is an integration constant given by initial
conditions. Now, these are also the equations for standard Einstein
gravity with an added cosmological constant, for a choice of
coordinates such that the metric determinant is equal to one, which is
always possible \cite{vanderBij:1981ym}. Therefore, the two theories
are classically equivalent, except that in the standard theory the
cosmological constant appears in the action, whereas in unimodular
gravity it is an integration constant. It has been shown
\cite{vanderBij:1981ym,Alvarez:2005iy,Unruh:1988in,Weinberg:1988cp}
that if one adds a matter sector that couples minimally to gravity,
and therefore has a covariantly conserved energy-momentum tensor
$\nabla_\mu T^{\mu\nu}=0$, the application of UG also results in the
appearance of an integration constant that plays the role of an
additional cosmological constant. We now want to find a similar
statement for a more general case, in particular the one in which
Newton's constant is generated dynamically. 

The action for unimodular gravity and any other fields, which couple
to gravity in an arbitrary way, has the following functional
dependence:
\begin{equation}\label{gphi}
\Sigma=
\int d^4x\mathcal{L}(\hat{g}_{\mu\nu},
\partial \hat{g}_{\mu\nu},\Phi,\partial\Phi),
\end{equation}
where $\Phi$ stands for all non-gravitational fields. If we want to
derive the equations of motion for this theory, we have to vary the
action keeping the constraint on the determinant. This is done using
the Lagrange multiplier method. We add an additional variable, whose
equation of motion will be the constraint. So, the following
Lagrangian is equivalent to the former one:
\begin{equation}
\mathcal{\tilde{L}}=\underbrace{\sqrt{-g}
\Big(\mathcal{L}(g_{\mu\nu},
\partial g_{\mu\nu},\Phi,\partial\Phi)+
\Lambda(x)\Big)}_{A}\underbrace{-\Lambda(x)}_{B}\;.
\end{equation}
Here, apart from the usual symmetry requirement
$g_{\mu\nu}=~g_{\nu\mu}$, $g_{\mu\nu}$ is unconstrained (the initial
Lagrangian was multiplied by a factor $\sqrt{-g}$, which does not
change the theory because of the unimodular constraint).

The equations of motion are
\begin{gather}
\frac{\delta A}{\delta g_{\mu\nu}}=0\;,\label{eqA}\\
\frac{\delta A}{\delta \Phi}=0\;,\\
\frac{\delta (A+B)}{\delta \Lambda}=0=(\sqrt{-g}-1)\;.
\end{gather}
We observe that $\int d^4 x\,A(x)$ is invariant under the full group
of diffeomorphisms. The infinitesimal transformations are
\begin{gather}
g_{\mu\nu}\rightarrow g_{\mu\nu}+\delta_\xi g_{\mu\nu}\;,\nonumber\\
\Phi\rightarrow \Phi +\delta_\xi \Phi\;,\nonumber\\
\Lambda\rightarrow\Lambda+\delta_\xi \Lambda\;,
\end{gather}
where $\delta_\xi$ depends on the nature of the fields, i.e. scalar,
vector, etc. If, for instance, we take $\Phi$ to be a scalar field,
the
$\delta_\xi$'s are given by
\begin{gather}
\delta_\xi g_{\mu\nu}=\nabla_\mu \xi_\nu +\nabla_\nu
\xi_\mu\;,\nonumber\\
\delta_\xi \Phi=\partial_\mu\Phi\xi^\mu\;,\nonumber\\
\delta_\xi \Lambda=\partial_\mu\Lambda\xi^\mu\;.
\end{gather}
Due to this symmetry, the following relation holds.
\begin{equation}
 \int d^4x\Big(\frac{\delta A}{\delta g_{\mu\nu}}
 \delta_\xi g_{\mu\nu}+\frac{\delta A}{\delta \Phi}
 \delta_\xi \Phi+\frac{\delta A}{\delta \Lambda}\delta_\xi
 \Lambda\Big)=0~.
\end{equation}
 The coefficients of the first two terms are zero because of the
equations of motion and the last coefficient yields $\frac{\delta
A}{\delta \Lambda}=\sqrt{-g}$. The equation reduces to
\begin{equation}
 \int d^4x\sqrt{-g}(\partial_\mu\Lambda)\xi^\mu=0\;.
\end{equation} 
Since this holds for all possible functions $\xi^\mu(x)$, we can
conclude that
\begin{equation}\label{constant}
 \partial_\mu\Lambda(x)=0\;,
\end{equation}
 and hence that $\Lambda$ is a constant of motion. Its value can be
determined by the field equations together with the initial conditions
for all fields. Knowing this, let us again look at the equations
\eqref{eqA}
\begin{equation}
 \frac{\delta A}{\delta g_{\mu\nu}}=
 \frac{\delta \{\sqrt{-g}
 \Big(\mathcal{L}(g_{\mu\nu},\partial g_{\mu\nu},\Phi,\partial\Phi)+
 \Lambda(x)\Big)\}}{\delta g_{\mu\nu}}=0\nonumber\;.
\end{equation}
These equations along with the constraint $\sqrt{-g}=1$ are the field
equations for unimodular gravity plus other fields. From
\eqref{constant} we know that $\Lambda$ is an integration constant. We
conclude that the theory given by \eqref{gphi} is
classically equivalent to a fully diffeomorphism invariant theory
described by the Lagrangian
\begin{equation}
\mathcal{L}_{\mathrm{diff}}= \sqrt{-g}
\Big(\mathcal{L}(g_{\mu\nu},\partial g_{\mu\nu},\Phi,\partial\Phi)
+\Lambda \Big)\;,
\end{equation}
apart from the different ways in which the parameter $\Lambda$
appears\footnote{In \cite{Buchmuller:1988wx} the authors presented a
proof of the same statement for TDiff theories a using similar type of
arguments. However, to our understanding, this proof is in fact  not
valid for general TDiff theories.}. The quantity $\Lambda$ plays the
role of a cosmological constant in the theory with explicit Planck
mass. However, as we will see shortly, this is not the case if
Newton's constant is induced dynamically.    

We now want to combine the ideas of UG and scale invariance. 
Considering only the gravitational and the scalar sectors, a general
Lagrangian containing scalar fields $\phi_i$ has the form:
\be
 \mathcal{L}=-\frac{1}{2}K_{ij}\phi_i\phi_j\hat{R}
 +\frac{1}{2}\hat{g}^{\mu\nu}\partial_\mu\phi_i\partial_\nu\phi_i
  -U_{ijkl}\phi_i\phi_j\phi_k\phi_l\;.
\ee
The result derived above tells us that the solutions of  UG with this
Lagrangian are equivalent to the solutions of GR with Lagrangian
\begin{equation}
\begin{split}
 \mathcal{L}=\sqrt{-g}\Big(-\frac{1}{2}K_{ij}\phi_i\phi_jR
 +\frac{1}{2}\sum_i g^{\mu\nu}\partial_\mu\phi_i\partial_\nu\phi_i\\
 - U_{ijkl}\phi_i\phi_j\phi_k\phi_l-\Lambda\Big)\;.
\label{gs} 
 \end{split}  
\end{equation} 
Let us finally add to unimodular gravity and scale invariance the
requirement that the scalar potential should have a flat direction.
The potential for a theory containing the Higgs field $h$ and an
additional scalar field $\chi$ is then given by
\begin{equation}
 V(h,\chi)= \frac{\lambda}{4}\left(h^2-\frac{\alpha}{\lambda}
\chi^2\right)^2\;.
\end{equation} 
So, our requirements lead us to the scalar and gravitational parts of
the action \eqref{lagr} with $\beta=0$ and standard gravity replaced
by UG. The corresponding  Lagrangian, invariant under all 
diffeomorphisms (a particular case of \eqref{gs}), is
\begin{equation}
\begin{split}
 \mathcal{L}&=\sqrt{-g}\Big(-\frac{1}{2}(\xi_{\chi}\chi^2
 +\xi_{h}h^2)R+\\
 &\frac{1}{2}
 \left(\partial_\mu \chi\right)^2 
 +\frac{1}{2}\left(\partial_\mu h\right)^2-V(h,\chi)
  -\Lambda\Big)\;.
\label{full} 
\end{split}
\end{equation}
Now, in order to facilitate the physical interpretation of this
Lagrangian, we can do a change of variables (conformal or Weyl
transformation) of the following type
\begin{equation}
 g_{\mu\nu}=\Omega(x)^2\tilde{g}_{\mu\nu}\;.
\end{equation} 
If we choose $\Omega$ such that $(\xi_{\chi}\chi^2 +\xi_{h}h^2)
\Omega^2=M_P^2$, the action \eqref{full} in terms of the new metric
$\tilde{g}_{\mu\nu}$ reads
\be
\mathcal{L}_{E} = \sqrt{-\tilde{g}}\left(
-M_P^2\frac{\tilde{R}}{2} + K
-U_E(h,\chi)\right)~,
\label{Eframe}
\ee
and is said to be in the Einstein frame (see e.g.
\cite{Faraoni:1998qx}). Here K is a complicated non-linear kinetic
term for the scalar fields, given by
\be
K=\Omega^2\left(\frac{1}{2}(\partial_\mu\chi)^2+
\frac{1}{2}(\partial_\mu
h)^2\right)- 3 M_P^2 (\partial_\mu\Omega)^2~.
\label{kin}
\ee
In our case where $\xi_\chi,\,\xi_h > 0$, the kinetic form $K$ is
positive-definite, which guarantees the absence of ghosts. The
Einstein-frame potential $U_E(h,\chi)$ is given by
\be
U_E(h,\chi)=\frac{M_P^4}{(\xi_{\chi}\chi^2
 +\xi_{h}h^2)^2}\left[V(h,\chi)+\Lambda\right]~,
\label{potE}
\ee
where the parameter $\Lambda$ is related to initial conditions for
scalar fields and gravity and does not depend on space-time
coordinates. It is not a cosmological constant but rather the strength
of a peculiar potential.

%%%%%%%%%%%%%%%%%%%%%%%%%%%%%%%%%%%%%%%%%%%%%%%%%%%%%%%%%%%%%%%%%%%%%%
\section{Dark energy, inflation and cosmological constant}
\label{sec:dark}
We now want to analyze the cosmological consequences of the theory
(\ref{lagr}), working in the Einstein frame. Although the dynamics of
the general system described by (\ref{Eframe}) is very complicated due
to the non-canonical form\footnote{For the system containing just one
scalar field the field transformation leading to a canonically
normalized kinetic term can be found easily  (see, e.g.
\cite{Fujii:2003pa}). For multiple fields we did not manage to find
the required transformation.} of the kinetic term $K$, we can gain
some insight into the evolution of the system looking at the potential
part only.

\begin{figure*}
\centerline{
\includegraphics*[width=5.5cm]{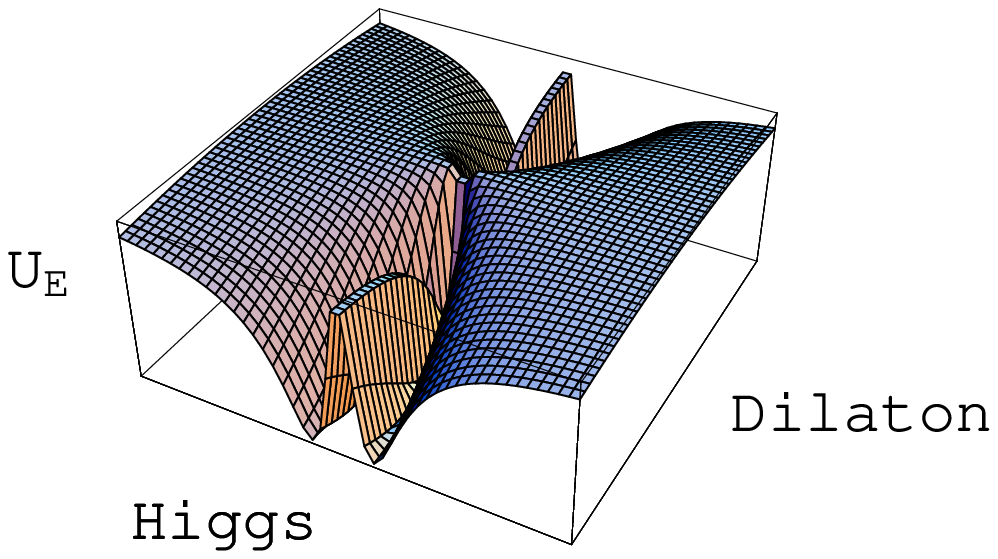}
\includegraphics*[width=5.5cm]{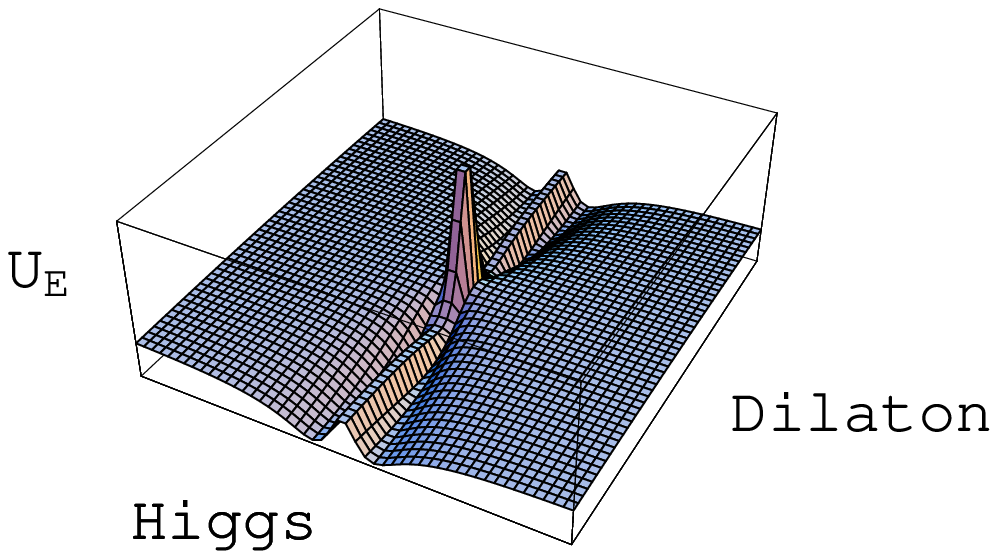}
\includegraphics*[width=5.5cm]{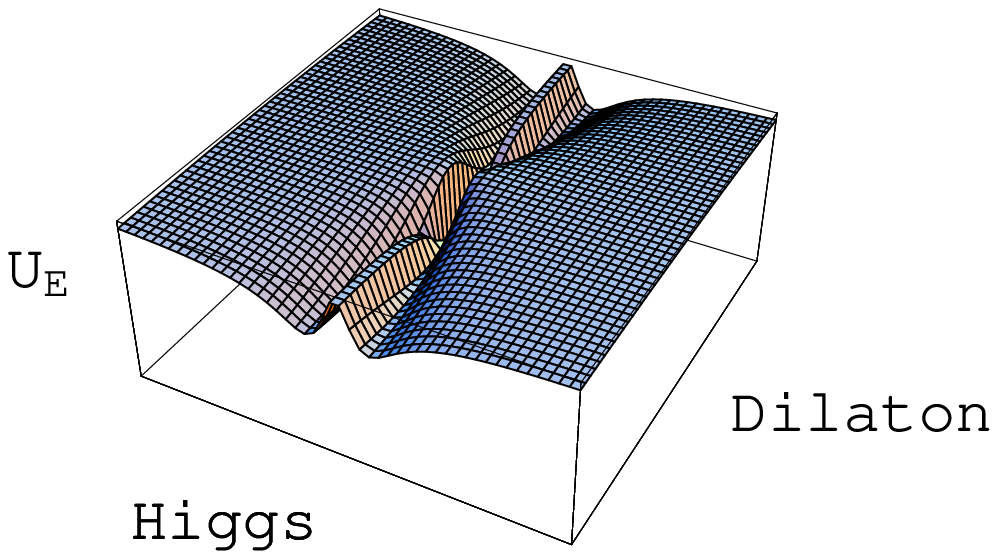}
}
\caption{Potential for the Higgs field and dilaton in the Einstein
frame. Left: $\Lambda=0$, middle: $\Lambda>0$, right $\Lambda<0$.} 
\label{fig:pot}
\end{figure*}

The phenomenologically interesting domain of parameters, explained
below, corresponds to ${\xi_h \gg 1}$, ${\xi_\chi \ll 1}$, ${\lambda
\sim 1}$, ${\alpha \simeq \lambda \xi_\chi \frac{v^2}{M_P^2}\lll 1}$
(here $v\simeq
250$ GeV is the Higgs vev). For this case the kinetic mixing of two
fields in $K$ is indeed not essential. The potential $U_E(h,\chi)$ for
$\Lambda=~0,~\Lambda>0$ and $\Lambda < 0$ is shown in Fig.
\ref{fig:pot}.
For $\Lambda=0$ it has two valleys around the lines  $h=\pm
\sqrt{\alpha/\lambda}\chi$, corresponding to the exact zero mode. As
soon as $\Lambda\neq 0$ the valleys get a slope. If $\Lambda > 0$, the
potential is positive for all field values and does not have any
minima but decreases if one moves away from the origin along a valley.
On the contrary, for $\Lambda<0$ the potential is negative for small
$\chi$ and $h$ and has a singularity at the origin. 

For $0<\Lambda\lsim {M_P^4}$ a typical behavior of the scalar fields
is as follows. Like in the chaotic inflation scenario
\cite{Linde:1983gd}, it is  expected that initially both fields are
displaced from their ground state values and are generically larger
than the Planck scale: the first term in (\ref{potE}) dominates.
Moreover, by assumption, $\xi_h \gg \xi_\chi$, meaning that for $\chi
\sim h$ the dynamics is mainly driven by the Higgs field, moving the
system towards the valley. This corresponds to inflation due to the
Higgs field, suggested in \cite{Bezrukov:2007ep}. When the value of
the Higgs field becomes of the order of the Planck scale, it is
trapped by the valley and oscillates there, producing particles of the
SM (this process is studied in detail in \cite{bgs}). The correct
spectrum of perturbations is generated if $\xi_h \sim 20000$
\cite{Bezrukov:2007ep}, for which the reheating temperature is $T_{rh}
\sim 10^{13}$ GeV. This part of the evolution is quite similar to the
hybrid inflation scenario \cite{Linde:1993cn}. The later evolution of
the universe depends crucially on the sign of $\Lambda$, which is
defined by initial conditions in UG. We would expect that with
$50\%$ probability  the universe was born in the state with
$\Lambda>0$. In this case it will evolve along the valley towards a
state with
$\chi,~h =\infty$ with zero cosmological constant. At any finite
evolution time the universe must contain dark energy. 

Present cosmological observations allow to pin down the value of the
non-minimal coupling of the field $\chi$. For the late time evolution
and $\alpha \ll 1$ ($\Rightarrow h\ll \chi$) the dilaton field
$\eta$
with (almost) canonical kinetic term is related to $\chi$ as
\be
\chi = M_P \exp\left(\frac{\gamma \eta}{4
M_P}\right),~~~\gamma=\frac{4}{\sqrt{6+\frac{1}{\xi_\chi}}}~.
\label{confchi}
\ee
Its dynamics is practically decoupled from the dynamics of the Higgs
field (the deviation of it from the vev  will be denoted by $\phi$ in
the Einstein frame). The corresponding equations of motion have the
form
\ba
\label{eqdil}
\ddot{\eta} + 3 H \dot{\eta} +\frac{d
U_\eta}{d\eta}=I_\eta~,\\
\ddot{\phi} + 3 H \dot{\phi} + m_h^2\phi=I_\phi~,
\label{eqhiggs}
\ea
where $m_h$ is the Higgs mass and
\be
U_\eta = \frac{\Lambda}{\xi_\chi^2} 
\exp \left(-\frac{\gamma \eta}{M_P}\right)~.
\label{potrho}
\ee
The source terms $I_{\eta,\phi}$ originate from the kinetic mixing
$K$ in eq. (\ref{kin}),
\ba
\nonumber
I_\eta &\propto&
\frac{1}{\xi_h}\left(\frac{\alpha\xi_h}{\lambda\xi_\chi}\right)^{\frac
{1}{2}}
\left(\ddot{\phi} + 3 H \dot{\phi}\right),\\
I_\phi &\propto&
\left(\frac{\alpha\xi_h}{\lambda\xi_\chi}\right)^{\frac{1}{2}}
\left(\ddot{\eta} + 3 H
\dot{\eta}-\frac{\gamma}{4M_P}\dot{\eta}^2\right)~,
\label{system}
\ea
and can be safely neglected. The Hubble constant $H$ is given by the
standard expression
\be
H^2 = \frac{1}{3M_P^2}\left(\frac{1}{2}\dot{\eta}^2 + U_\eta +
\frac{C_\gamma}{a^4}+ \frac{C_M}{a^3}\right)~,
\ee
where the two last terms correspond to radiation and matter
contributions to the energy density, and $a$ is the scale factor. It
is amazing that here the exponential potential, proposed for a
quintessence field a long time ago in 
\cite{Wetterich:1987fk,Wetterich:1987fm,Ratra:1987rm}, appears
automatically, though with $\Lambda$ not being a fundamental parameter
but rather a random initial condition.

The dynamics of the universe described by eq. (\ref{eqdil}) with
$I_\eta=0$ has been studied in a number of works (for a recent review
see \cite{Copeland:2006wr}). In
\cite{Wetterich:1987fm,Ferreira:1997hj} it was shown
that for $\gamma>\sqrt{3}$ this model possesses attractors
corresponding to scaling solutions. In that case the energy density of
the scalar field eventually scales like the dominant component of the
universe. Therefore, those models cannot describe accelerated
expansion. The situation is different for $\gamma<\sqrt{3}$: the
scalar component changes slower than radiation and matter and
eventually starts dominating. If $\gamma$ is in this region, the
dynamics of $\eta$ is that of a ''thawing'' quintessence field
\cite{Caldwell:2005tm,Ferreira:1997hj}. In that scenario the scalar
field at early times is nearly constant and has $\omega\simeq -1$.
When the Hubble friction gets weaker, the field starts rolling down
the exponential potential. At the same time $\omega$ grows and moves
away, although extremely slowly, from $\omega=-1$.

Let us find a constraint on the parameter $\xi_\chi$. The most
convenient for us is the result of \cite{Scherrer:2007pu}, which gives
the relationship between the parameter $\omega$ of the equation of
% state for dark energy $p=\omega\rho$ and the dark energy abundance
$\Omega_\chi$, valid for the thawing scenario realized by an
exponential potential with $\gamma<\sqrt{3}$. It reads
\cite{Scherrer:2007pu}:
\be
1+\omega = \frac{\gamma^2}{3}\left[\frac{1}{\sqrt{\Omega_\chi}}
-\frac{1}{2}\left(\frac{1}{\Omega_\chi}-1\right) 
\log\frac{1+\sqrt{\Omega_\chi}}{1-\sqrt{\Omega_\chi}}\right]^2~.
\label{eos}
\ee
Taking the results of WMAP \cite{Komatsu:2008hk} for the equation of
state ${-0.14<1+\omega<0.12}$ and identifying $\Omega_\chi$ with the
dark energy abundance ${\Omega_{DM}\simeq 0.73}$, we find that the
value of $\xi_\chi$ must be in
the interval
\be
0< \xi_\chi< 0.07~.
\label{chiint}
\ee
Hence, $\gamma\lesssim 1<\sqrt{3}$ which shows that $\gamma$ is
indeed in the correct parameter region for the thawing scenario.

As for the value of $\Lambda$, it cannot be determined unambiguously,
since the value of the dilaton field is unknown.  For typical values
appearing in run-away scenarios,  $\eta \sim M_P
\log(t M_P)/2\sqrt{\xi_\chi}
$ \cite{Wetterich:1987fk,Wetterich:1987fm} ($t$ is the
present age of the universe), the initial value of $\Lambda$ can be as
large as $M_P^4$. 

It has been shown in \cite{Wetterich:1987fm,Ferreira:1997hj}, that in
this model (for
$\Lambda>0$) the universe becomes dark energy dominated at late times,
i.e. $\Omega_{\chi}\rightarrow 1$. In this limit the parameter of the
equation of state becomes $\omega\rightarrow\frac{\gamma^2}{3}-1$ and
the universe will expand according to the power law
$a(t)\propto t^{2/\gamma^2}$.

An important comment is now in order. The change of $\eta$ with time
{\em does not lead} to any visible time variation of Newton's constant
or of particle masses. In the Einstein frame the dilaton practically
decouples from all the fields of the SM, and the amplitude of the
time-dependent corrections to masses, from eq. (\ref{eqhiggs}), is of
the order of 
\be
\frac{\phi}{v} \sim \frac{I_\phi}{m_h^2 v}\;,
\label{deviate}
\ee 
which is too small to be tested in any observations. 

To conclude, the cosmological evolution of a {\em classical}
scale-invariant theory with unimodular gravity and an exactly massless
dilaton typically leads to initial inflation (large $\xi_h \gg 1$ is
required to be in accordance with observations), then to the heating
of the universe and eventually, with a $50\%$ chance, to accelerated
expansion in the late stages ($\xi_\chi\ll 1 $ is required to fit
observations). 

%%%%%%%%%%%%%%%%%%%%%%%%%%%%%%%%%%%%%%%%%%%%%%%%%%%%%%%%%%%%%%%%%%%%%%
\section{Quantum theory}
\label{sec:dreams}
The analysis of the two previous sections was entirely classical.
Therefore, we will formulate the conditions, which should be satisfied
for the results to be valid in the quantum case as well.  Since the
theory (\ref{full}) is not renormalizable, the discussion in this
section will be on the level of wishful thinking and does not pretend
to any rigor (see, however,
\cite{Shaposhnikov:2008xi,Shaposhnikov:2008ar}). 

As the dilatational symmetry of the theory is easier to see in the
Jordan frame, we will use it for the present discussion, which in a
number of  respects resembles the one in \cite{Wetterich:1987fk,
Wetterich:1987fm, Wetterich:2008sx}. Clearly, the classical results
survive if the dilatational symmetry remains exact on the quantum
level
and if the dilaton is still massless in full quantum field
theory. Like in the classical case, the exact dilaton degeneracy of
the ground state will guarantee the absence of the cosmological
constant, whereas the unimodular character of
gravity would induce, through initial conditions, the run-away
behavior (for $\Lambda >0$) of the dilaton field at a late time in
the expansion of the universe. If true, {\em all} dimensionful
parameters of the SM, including those coming from dimensional
transmutation like $\Lambda_{QCD}$ will change in the same way during
the run-away of the dilaton field. The deviation of dimensionless
ratios (only they are relevant for physics in a scale-invariant
theory) from constants, due to the cosmological evolution, will be
strongly suppressed as in (\ref{deviate}) and thus be invisible. The
dilaton will only have derivative couplings to the fields of the
$\nu$MSM, being a Goldstone boson related to the spontaneous breaking
of dilatational invariance and thus evade all the constraints
\cite{Wetterich:1987fk, Wetterich:1987fm, Wetterich:2008sx} considered
for the Brans-Dicke field \cite{Brans:1961sx}. 

The required dependence of low-energy parameters such as
$\Lambda_{QCD}$ on the dilaton field would appear if the following
strategy is applied to the computation of radiative corrections in the
constant $\chi,~h$ backgrounds
\cite{Shaposhnikov:2008xi,Shaposhnikov:2008ar} (see also
\cite{Wetterich:1987fk, Wetterich:1987fm}). Use the field-dependent
cutoff $Q^2$, related to the effective Planck scale in the Jordan
frame as $Q^2 = \xi_\chi \chi^2 + \xi_h h^2$, and assume that the
values of all dimensionless couplings at this scale do not depend on
$Q^2$. With this prescription the non-trivial dimensionful
parameters, appearing as a result of the renormalization procedure,
would acquire the necessary dependence on the dilaton field. It is
this prescription, which was effectively used in
\cite{Bezrukov:2007ep} to discuss radiative corrections to the
Higgs-inflaton potential. Note that the renormalization procedure of 
\cite{Barvinsky:2008ia} is completely different, and there is no
surprise that the results and conclusions of our present work and of 
\cite{Bezrukov:2007ep} are not the same as  those of
\cite{Barvinsky:2008ia}.

The requirement that the dilaton remains exactly massless on the
quantum level, or in other words that the quantum effective potential
has a flat direction (in classical theory this corresponds to
$\beta=0$) is crucial for our findings. It is highly non-trivial as it
is not a consequence of scale symmetry. Still, the quantum field
theories constructed near the scale-invariant ground state
$\langle h\rangle=\langle\chi\rangle=0$
and the state with spontaneously broken scale invariance are
completely different at the quantitative level. The theory with
$\langle h\rangle=\langle\chi\rangle=0$ in general does not have
asymptotic scattering
states\footnote{\label{ft}If it does, the exact propagators coincide
with the free ones and the theory is  likely to be trivial in this
case \cite{Greenberg:1961mr,Bogolyubov:1975ps}.}  (for a review see
e.g. \cite{Aharony:1999ti}). In other words, it does not have
particles at all and thus  cannot be accepted as a realistic theory. 
On the contrary, if the quantum scale invariance is {\em
spontaneously} broken, the theory does describe particles and thus may
be  phenomenologically relevant. These considerations single out the
theories with degenerate ground
state, corresponding to $\beta=0$ at the classical level\footnote{When
quantum effects are included, the flat direction is not necessary
associated with $\beta=0$. Nevertheless, to simplify the discussion,
we will refer to the case with spontaneously broken scale invariance
on the quantum level still quoting the classical value $\beta=0$.}.

In the cosmological setup the relevance of these arguments is
not  obvious. Indeed, suppose that the flat direction is lifted, i.e.
$\beta >0$ and consider late time classical evolution corresponding to
initial conditions with $\Lambda >0$. A non-zero value of $\beta$
leads
to a positive vacuum energy in the Einstein frame, $E_{vac}\sim\beta
M_P^4$. As in Section \ref{sec:dark}, the scalar field $\eta$ has a
run-away behavior, leading to the breaking of scale invariance,
whereas the universe is expanding exponentially with the Hubble
constant determined by $E_{vac}$. Certainly, there is nothing wrong
with the classical solution of this type, except the fact that it is
unstable against scalar field fluctuations which grow as in the
inflationary stage \cite{Mukhanov:1981xt}.  Whether this  picture
survives quantum-mechanically, is an open issue. If it does, and if
the
notion of particles can be defined, we will get a universe with
non-zero cosmological constant, loosing thus an explanation of its
absence. However, the classical picture may happen to be misleading.
Indeed, in the full quantum field theory, any state including
cosmological solutions, can be considered as a superposition of
quantum excitations above the vacuum state. As no particles can be
defined for the scale-invariant theory with scale-invariant ground
state\footnote{The case when particles can be defined corresponds to
free field theory, see  footnote \ref{ft}. Then the Lagrangian 
(\ref{lagr}) is simply a very complicated way to describe this trivial
theory, and the classical cosmological solution has nothing to do with
the exact quantum solution.}, we would expect that the effective
particle-like excitations about the classical background can exist
only for a finite time, presumably of the order of the inverse Hubble
constant. If true, the requirement that the scale-invariant quantum
theory must be able to describe particles again singles out the case
corresponding to $\beta=0$.   

%%%%%%%%%%%%%%%%%%%%%%%%%%%%%%%%%%%%%%%%%%%%%%%%%%%%%%%%%%%%%%%%%%%%%%
\section{Conclusions}
\label{sec:concl}
In this Letter we showed that the scale-invariant classical SM and the
$\nu$MSM coupled to unimodular gravity have all necessary ingredients
to describe the evolution of the universe, including early inflation
and late acceleration. The requirement of scale invariance and of the
existence of a massless dilaton leads to a theory in which all mass
scales, including that of gravity, originate from one and the same
source. The unimodular character of gravity leads to the generation of
an exponential potential for the dilaton, ensuring the existence of
dark energy. If the full quantum field theory exhibits the same
symmetries (for the explicit construction see
\cite{Shaposhnikov:2008xi,Shaposhnikov:2008ar}), it will have the same
properties. Moreover, the argument can be reverted -- the observation
of an accelerated universe with a dark energy component may tell that
the underlying quantum field theory should be scale-invariant, and
that this scale invariance should be broken spontaneously, leading to
the massless dilaton.

The theory (\ref{lagr}) with $\beta=0$ and UG happens to be very
rich in applications. It can address all confirmed signals which
suggest that the SM is not complete: neutrino masses and oscillations,
existence of dark and baryonic matter in the universe, early inflation
and late acceleration, leading to an extra argument  against the
necessity of new physics between the electroweak and  the Planck scale
\cite{Shaposhnikov:2007nj} (see also \cite{Meissner:2006zh}). The
discussion of particle physics experiments and astrophysical
observations that can confirm or rule out this theory can be found in
\cite{Boyarsky:2006fg,Bezrukov:2006cy,
Bezrukov:2005mx,Gorbunov:2007ak,Boyarsky:2006hr}. Our findings here
indicate that the equation of state parameter $\omega$ for dark energy
must be different from that of the cosmological constant, but also
that $\omega > -1$, adding an extra cosmological test that could rule
out the $\nu$MSM.  

%%%%%%%%%%%%%%%%%%%%%%%%%%%%%%%%%%%%%%%%%%%%%%%%%%%%%%%%%%%%%%%%%%%%%%
{\bf Acknowledgements.}
This work was supported by the Swiss National Science Foundation. We
thank F. Bezrukov, K. Chetyrkin, S. Sibiryakov and I. Tkachev for
valuable comments.

\end{document}